\documentclass{revtex4}

\usepackage{amsmath}
\usepackage{enumerate}
\newcommand{\assign}[1]{}

\usepackage{amssymb,graphicx}
\usepackage{epsfig}
\usepackage{booktabs}
\usepackage[usenames]{color}
\usepackage{multirow}
\usepackage{rotating}

\newcommand{\had}{{\sc had\_amr}}

\newcommand{\I}[1]{\textit{#1}}
\newcommand{\B}[1]{\textbf{#1}}

\begin{document}

\title[An Application Driven Analysis of the ParalleX Execution Model]
{An Application Driven Analysis of the ParalleX Execution Model}

\author{Matthew Anderson$^{1}$,
        Maciej Brodowicz$^{1}$,
        Hartmut Kaiser$^{1,2}$,
        Thomas Sterling$^{3}$}
\affiliation{$^{1}$Center for Computation and Technology, Louisiana State University, Baton Rouge, LA, \\
             $^{2}$Department of Computer Science, Louisiana State University, Baton Rouge, LA, \\
             $^{3}$Center for Research in Extreme Scale Technology, Indiana University, Bloomington, IN}
\email{matt@phys.lsu.edu, \{maciek, hkaiser\}@cct.lsu.edu,sterling@iu.edu}

%\category{C.2}{Computer Communication Networks}{Computer Networks}
%\category{C.4}{Performance of Systems}{Analytical Models}
%\category{G.3}{Stochastic Processes}{Queueing Systems}
%\terms{Internet Technologies, E-Commerce}
\keywords{Execution Model; ParalleX; Model of Computation; Parallel Processing; 
          high performance computing; adaptive mesh refinement}

\begin{abstract}
Exascale systems, expected to emerge by the end of the next decade, 
  will require the exploitation of
  billion-way parallelism at multiple hierarchical levels in order to
  achieve the desired sustained performance.
  The task of assessing future machine performance is approached 
  by identifying the factors which
  currently challenge the scalability of parallel applications.
  It is suggested that the root cause of these challenges is the
  incoherent coupling between the current enabling technologies,
  such as Non-Uniform Memory Access of present multicore 
  nodes equipped with optional hardware accelerators and
  the decades older execution model, i.e., the Communicating Sequential
  Processes (CSP) model best exemplified by the message passing interface (MPI) 
application programming interface. 
  A new execution model, ParalleX, is introduced as an alternative to the CSP model.
  In this paper, an overview 
  of the ParalleX execution model is presented along with details about
  a ParalleX-compliant runtime system implementation called High Performance ParalleX (HPX).  
  Scaling and performance results for an adaptive mesh refinement numerical 
  relativity application developed using HPX are discussed.   The performance results 
  of this HPX-based application are compared with a counterpart MPI-based mesh refinement
  code.  The overheads associated with HPX are explored and
  hardware solutions are introduced for accelerating the runtime system.
\end{abstract}

\maketitle

\section{Introduction}
An entire class of parallel applications is emerging that is
scaling-impaired. These are simulations that consume extensive
execution time, sometimes exceeding a month, but which are not able to
use effectively more than a few hundred processors.  These applications require a dramatic
reduction of execution time for fixed workloads but suffer from poor strong scaling behavior.
One such class of applications is based on Adaptive Mesh Refinement (AMR) algorithms which
 concentrate processing effort at the most dynamic parts of the computation domain. 
AMR is employed in many applications from astrophysics and numerical relativity to
Navier-Stokes solvers.  Today's conventional parallel programming methods such as 
MPI~\cite{MPISpec} and systems such as distributed memory massively parallel processors (MPPs) and Linux clusters 
exhibit poor efficiency and constrained scalability for this class of applications. 
 This severely hinders scientific advancement.  
Many other classes of applications exhibit similar properties, especially graph/tree 
data structures that have non uniform data access patterns.

An underlying hypothesis of this work is that achieving the goal of
dramatic scalability improvements for both current strong scaling impaired
applications and future Exascale applications will require a new execution model to
replace the conventional communicating sequential processes (CSP) model
best exemplified by the MPI application programming interface. It is
noted that this position is controversial and a focus of community-wide
debate.  The ExaScale computing study~\cite{ExaScale_study} concluded that a new execution model and 
programming methodology is required for dramatic scalability improvements in such problems.
This paper briefly presents such a model, ParalleX, and provides early results 
from an experimental implementation of an AMR application exploring the threshold of 
singularity formation and critical behavior in numerical relativity.  
This AMR application is based on the prototype HPX runtime system which is an early 
implementation of the ParalleX model used as an experimental framework.

This work is motivated by the dual challenge of
applications which through conventional practices either are presently
unable to effectively exploit a relatively small number of cores in a
multi-core system or that by the end of this decade will not be able to
exploit Exascale computing systems likely to employ hundreds of millions
of such cores. We consider four factors inhibiting
these two forms of scalability: 1) \I{starvation} that is the insufficiency
of availability of useful work either globally or locally, 2) \I{latency}
that is the distance measured in time (e.g., cycles) for a remote access
or service request, 3) \I{overhead} that is the critical time and work required
to manage parallel resources and concurrent tasks which would not be
required for pure sequential execution, and 4) waiting for \I{contention} or
delays due to conflicts for shared physical or logical resources. 

  The ParalleX execution model~\cite{scaling_impaired_apps} has been developed to address these challenges by 
enabling a new way of computation based on message-driven flow control in a 
global address space coordinated by lightweight synchronization semantic constructs.
The key features of ParalleX that provide significant advantages over the CSP model
are message driven computation based on parcels, split phase transaction, 
light weight synchronization using local control objects including futures and dataflow,
and fine grain multithreading.
In the following section we will describe the ParalleX model and introduce the prototype
HPX runtime system that delivers the mechanisms required to support the parallel execution, 
synchronization, resource allocation, and name space management. 

\section{The ParalleX Execution Model}
\label{sec:hpx}

An execution model is a set of governing principles
that guide the co-design, function, and interoperability of all layers
of the system's structure from the programming model through the
system software to the hardware architecture.  Included among such 
principles are the system semantics, referentiable structures, 
naming, communication, parallel control paradigm including synchronization, 
and policies of resource management.

ParalleX is an experimental execution model to serve as a framework 
for research into science application scalability and future 
high performance computing (HPC) system hardware and software design and
operation.
As noted in the introduction, ParalleX is motivated by: (1) the long term objective of enabling Exaflops
scale computing by the end of the decade in an era of flattening clock
rates and processor core design complexity resulting in the expected
integration of up to a billion cores by the beginning of the next
decade; (2) the more immediate scaling concerns of a diverse set of
what we refer to as \textit{scaling-challenged} problems that do not
scale well beyond a small number of cores, and take a long time to complete.
Key aims of ParalleX include:
\begin{itemize}
\item expose new forms of program parallelism including fine grain
  parallelism to increase the total amount of concurrent operation;
\item reduce overhead for efficiency of operation and, in particular, to
  make effective use of fine grain parallelism where it should
  occur; this includes, where possible, the elimination of global barriers;
\item facilitate the use of dynamic methods of resource management and
  task scheduling to exploit runtime information about the execution
  state of the application and permit continuing adaptive control for
  best causal operation.
\end{itemize}
ParalleX, like any true model of computation, transcends any single element of a high 
performance computer to represent the holistic system structure, interdependencies, 
and cooperative operation of all system component layers. 

While ParalleX incorporates many useful concepts developed
elsewhere, some extending back as much as three decades, it
constitutes a new synthesis of these as well as innovative ideas in a
novel schema that is distinct from conventional practices and that
exhibits the necessary properties identified above to increase
application and system scalability.
The form and function of the current ParalleX model consist of six key concepts
or management principles: ParalleX Processes, the Active Global Address Space (AGAS), threads and
their management, parcel transport and parcel management, Local Control Objects (LCOs),
and percolation.  With the exception of processes and percolation, all have been incorporated
in a C++ prototype runtime implementation of ParalleX called HPX.   Each concept is described below
along with brief details about the HPX implementation:

\B{AGAS -- The Active Global Address Space:} The requirements for dynamic
load-balancing and the support for dynamic AMR related problems define the
necessity for a single global address space across the system. This not only
simplifies application writing, as it removes the dependency of codes on static
data distribution, but enables seamless load-balancing of application and system
data. An active global address space overcomes the disadvantages of prior systems,
such as X10~\cite{Charles:2005:XOA:1103845.1094852}, Chapel~\cite{Chamberlain07parallelprogrammability},
or UPC~\cite{UPCSpec}, as it allows for fully dynamic adaptive resource management.
The abstraction of localities is introduced as a means of defining a
border between controlled synchronous (intra-locality) and fully asynchronous
(inter-locality) operations. A locality is a contiguous physical domain,
managing intra-locality latencies, while guaranteeing compound atomic operations
on local state. Different localities may expose entirely different temporal
locality properties. Our implementation interprets a locality to be equivalent
to a (cluster-) node in a conventional system. Intra-locality data access means access to
the local memory, while inter-locality data access and data movement
depend on the system network. In ParalleX, referencing first class objects, such as threads,
processes, or Local Control Objects (LCOs),  is decoupled from its locality.

\B{Threads and their Management:} The HPX-thread manager implements a work queue
based execution model very similar to prior systems (Cilk++~\cite{cilk++},
TBB~\cite{inteltbb}, PPL~\cite{microsoftppl}).  In addition, HPX-threads are first
class objects with immutable global names, enabling even remote management.
We avoid moving threads across localities (expensive operation); instead, work
migrates via \I{continuations} by sending a \I{parcel} that might cause the instantiation
of a thread at the remote locality.
The difference between moving a thread across a locality and migrating work
via \I{continuations} is one of complexity: moving a thread is much more complex.  
Moving a thread across localities includes moving both the context (stack frame) 
and registers.  A continuation involves just the locality identifier and arguments.
 HPX-threads are cooperatively (non-preemptively)
scheduled in user mode by a thread manager on top of a static OS-thread per core. The
HPX-threads can be scheduled without a kernel transition, which provides a
performance boost. Additionally the full use of the OS's time quantum per
OS-thread can be achieved even if an HPX-thread blocks for any reason.

\B{Parcel Transport and Parcel Management:} In ParalleX, \I{parcels} are an extended
form of active messages~\cite{Wall:1982:MAA:582153.582157} for inter-locality communication.
Parcels are the remote
semantic equivalent to creating a local HPX-thread. If a function is to be
applied locally, an HPX-thread is created; if it has to be applied remotely, a
parcel is generated and sent which will create an HPX-thread at the remote site.
Parcels are either used to move the work to the data (by applying an operation
on a remote entity) or to gather small pieces of data back to the caller.
Parcels enable the message driven paradigm (as developed in TAM~\cite{Culler93acompiler},
Split-C~\cite{citeulike:7729323}) for distributed control flow and for dynamic
resource management, featuring a split-phase transaction based execution model.
While the current HPX implementation of ParalleX relies on TCP/IP, 
work is in progress to move to high performance messaging libraries, such 
as GASNet~\cite{gasnetspec} and
Converse~\cite{converse}.

\begin{figure} %[][htp] %{l}{0.6\textwidth}
  \includegraphics[width=0.99\linewidth]{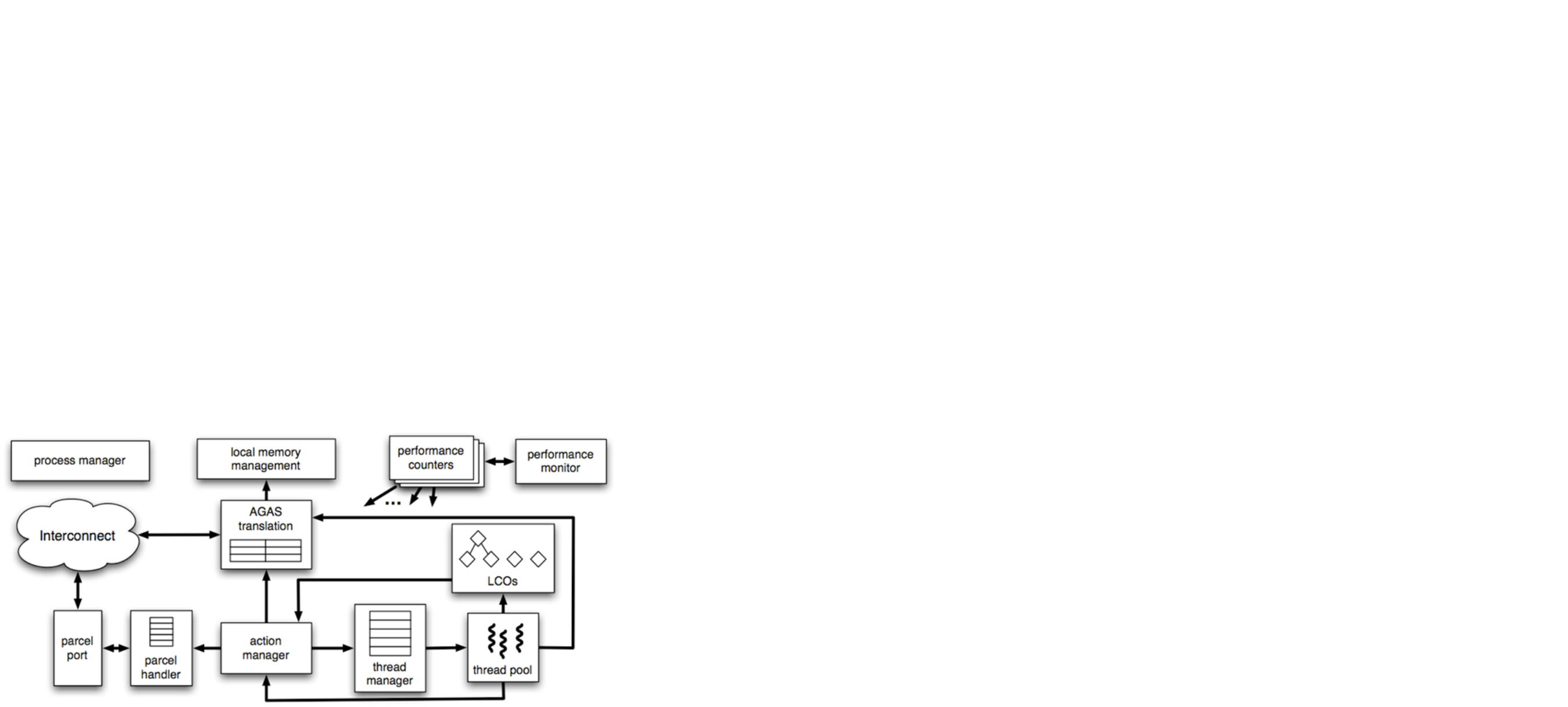}
  \caption{\small{Modular structure of HPX implementation. HPX implements the supporting
     functionality for all of the elements needed for the ParalleX
     model: AGAS (active global address space), parcel port and
     parcel handlers, HPX-threads and thread manager, ParalleX processes, LCOs (local control objects),
     performance counters enabling dynamic and intrinsic system and load
    estimates, and the means of integrating application specific components.}
  }
\label{fig:hpxarch}
\end{figure}

\B{Local Control Objects (LCOs):} An LCO is a synchronization abstraction of different
functionalities for event-driven HPX-thread creation, protection of data
structures from race conditions and automatic event driven on-the-fly scheduling
of work with the goal of letting every single function proceed as far as
possible. 
LCOs are used to
organize flow control. A well known and prominent example of an LCO is a
\I{future}~\cite{futures,  Halstead:1985:MLC:4472.4478, Baker:1977:IGC:872736.806932}.
It refers to an object that acts as a proxy for a result that is initially not
known, usually because the computation of its value has not yet completed. The
\I{future} synchronizes the access to this value by optionally suspending
the requesting thread until the value is available. This allows the computation
to proceed unblocked until the actual value is required to produce a result rather
than, say, incorporating it into a more complex data structure.  Futures also
permit anonymous producer-consumer computation when neither the producer
of a value, nor its consumer are known at compile time. In addition, the future
construct allows a tradeoff between eager and lazy evaluation by postponing
the calculation of a value until it is actually required.
Another example is the \I{dataflow} LCO.
It defines the events or precedence that must be satisfied in
order to perform a follow-on action (e.g., thread). Named after the
early experimental execution model of the 1970s and 1980s~\cite{Dennis74,DennisM98,dyn_dataflow},
dataflow LCOs provide a powerful semantic mechanism for managing asynchrony of
operation while yielding lightweight control to eliminate (in most
cases) the use of global barriers. The dataflow LCO construct acquires
result values (or references) and is event driven updating its internal
state accordingly until one or more precedent constraints are satisfied;
then it initiates further program action dependent on this/these
conditions. Not only
does this automatically allow computation and communication to overlap
(thus hiding latencies given sufficient parallelism) but also allows many
phases of the computation to overlap thereby exposing more parallelism
at a given time.

HPX provides specialized
implementations of a full set of synchronization primitives (futures, dataflow LCOs, mutexes,
conditions, semaphores, full-empty bits, etc.) usable to cooperatively block a
HPX-thread while informing the thread manager that other work can be run on the
OS-thread (core). The thread manager can then make a scheduling decision to
execute other work.

\B{ParalleX Processes:} ParalleX establishes a new relationship between virtual
processes and the physical processing resources. Conventional practices assign
a given process to a specified processor (or core). ``Parallel processes'' means
multiple processes operating concurrently. ParalleX parallel processes
incorporate substantial parallelism and map to multiple
cores. A ParalleX parallel process provides part of the global name space for its internal
active entities, which include other localities, child-processes, threads, data, methods, and 
physical allocation mappings. It allows
application modules to be defined with a shared name space and to exploit many
layers of parallelism within the same context.
Processes are ephemeral, being instantiated during runtime and
exhibiting finite life cycle at the conclusion of which they are terminated.
As mentioned, the HPX implementation of ParalleX does not support this currently.    
HPX is the first implementation of ParalleX and has limited functionality.  Processes
are needed for distributed data, locality control, task instantiation, and policy 
management.    

Among the key features of the HPX C++ implementation of ParalleX (See Fig.~\ref{fig:hpxarch}), we note that:
\begin{itemize}
\setlength{\itemsep}{0pt}
  \setlength{\parskip}{0pt}
  \setlength{\parsep}{0pt}
  \item it is a modular, feature-complete, and performance oriented representation of the ParalleX model targeted at conventional architectures 
           and, currently, Linux based systems, such as symmetric multiprocessing (SMP) nodes and conventional clusters; 
  \item it has a modular architecture which allows for easy compile time customization and minimizes the runtime memory footprint;  
  \item it enables dynamically loaded application-specific modules to extend the available functionality at runtime (static pre-binding at link time is also supported);
  \item it strictly adheres to Standard-C++~\cite{cpp_standard} and utilizes Boost~\cite{boost}, 
enabling it to combine powerful compile time optimization techniques and optimal code generation with excellent portability.
\end{itemize}

A walkthrough description of the HPX architecture is found in Figure~\ref{fig:hpxarch}.  An incoming parcel (delivered over the interconnect) is received by the parcel port. One or more 
parcel handlers are connected to a single parcel port, optionally allowing to distinguish different 
parts of the system as the parcel's final destination. An example for such different destinations 
is to have both normal cores and special hardware (such as a GPGPU) in the same node. The 
main task of the parcel handler is to buffer incoming parcels for the action manager. The action 
manager decodes the parcel and creates a PX-thread based on the encoded information. 
All PX-threads are managed by the thread manager, which schedules their execution on one 
of the OS-threads. Usually HPX creates one OS-thread for each available core. The thread 
manager has implemented several scheduling policies, such as a global queue scheduler, where all
cores pull their work from a single, global queue, or a local priority scheduler, where each core
pulls its work from a separate priority queue. The latter supports work stealing for better
load balancing.

If a possibly remote action has to be executed by one of the PX-threads, the action manager 
queries AGAS whether the target of the action is local or remote to the node the PX-thread 
is running on. If the target happens to be local, a new PX-thread is created and passed to the 
thread manager. This thread encapsulates the work (function) and the corresponding arguments 
for that action. If the target is remote, the action manager creates a parcel encoding the 
action (i.e. the function and its arguments). This parcel is handed to the parcel handler, which 
makes sure that it gets sent over the interconnect.

The PX processes, the local memory management, the performance counters (a generic monitoring 
framework), and the LCOs are all implemented on top of an underlying component framework. 
Components are the main building blocks of remotable actions and could encapsulate arbitrary, 
possibly application specific functionality. In the case of the mentioned components,
the HPX runtime system implements its own functionality in terms of this component framework.
Typically any application written using HPX extends the set of existing components based on
its functionality requirements.

\section{AMR-based application}
\label{sec:amr}

Modern finite difference based simulations require adequately
resolving many physical scales which often vary over several orders of
magnitude in the computational domain.  Many high performance
computing toolkits have been developed to address this need by
providing distributed AMR based on the MPI libraries \cite{Carpet, SAMRAI,had_webpage,Chombo,PAMR,Paramesh,AMROC}.  
The AMR algorithm, introduced by Berger-Oliger~\cite{Berger}, employs multiple computational grids of 
different resolution and places finer-resolution meshes where needed in the computational domain in order to 
adequately resolve phenomena at increasingly smaller physical and temporal scales.  

\begin{figure} \centering
\epsfig{file=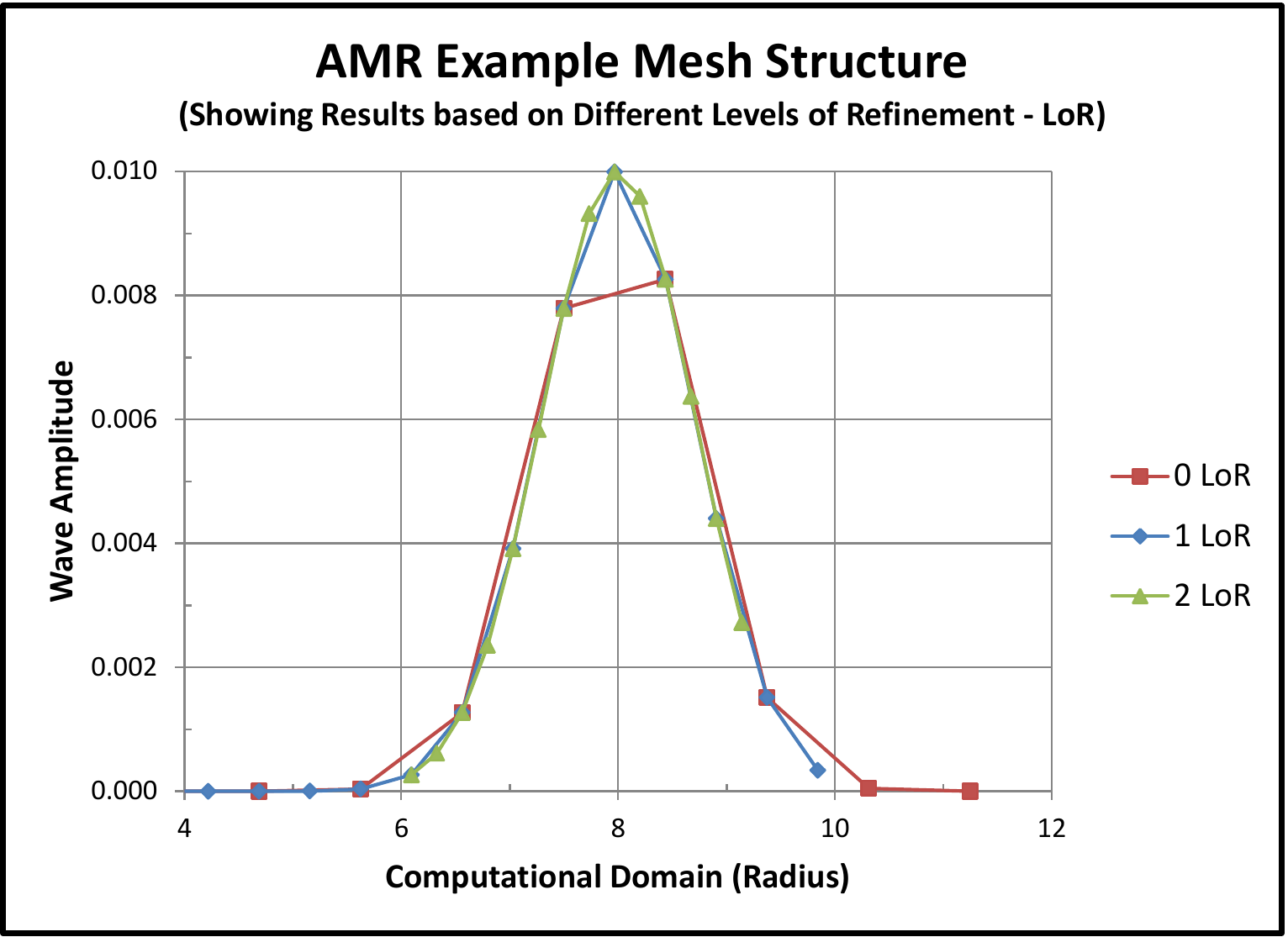, width=0.95\linewidth} %,height=10.0cm,angle=270}
\caption{\small{Two levels of AMR (three different resolution meshes) at the initial timestep of a time-dependent simulation.  Independent variables are time and radius; wave amplitude is a dependent variable.  More resolution is placed
where truncation error is highest.}
}
\label{fig:amr_example}
\end{figure}

3-D AMR simulations are typically $10^4$--$10^5$ times faster than performing a computation using a single 
resolution mesh.  A sample initial AMR mesh structure of the test application we explore here is 
illustrated in Fig.~\ref{fig:amr_example}.
The initial data supplied is a wave pulse.  As the wave pulse moves, the 
higher resolution meshes adjust accordingly in order to keep the local error criterion below
threshold.  In discussing AMR scaling, we differentiate scaling characteristics into two types: strong and weak scaling.  
In strong scaling, the test application problem size is kept constant while the number of processors 
devoted to computing the simulation is increased.  In weak scaling, the problem size is increased as the 
number of processors is increased so that the local processor workload on the system is kept constant.

The application is a nonlinear wave equation in spherical symmetry from critical phenomena~\cite{liebling}:
\begin{eqnarray}
\dot{\chi} &=& \Pi \label{eq:chi}  \\
\dot{\Phi} &=& \frac{\partial{\Pi}}{\partial{r}} \\
\dot{\Pi} &=& \frac{1}{r^2} \frac{\partial{\left(r^2 \Phi\right)}}{\partial r} + \chi^p 
\label{eq:critical}
\end{eqnarray}
where $p=7$.
Second order finite differencing is used in space and the system is integrated in time
using Runge-Kutta third order.  The initial data are  
\begin{eqnarray*}
\chi_0 &=& A \exp{\left[-(r-R_0)^2/\delta^2\right] } \\ 
\Phi_0 &=& \frac{\partial \chi_0}{\partial r} \\
\Pi_0 &=& 0
\end{eqnarray*}
where parameters $R_0=8$, $\delta=1$, and the amplitude $A$ is tuned to explore criticality.
The AMR algorithm is Berger-Oliger~\cite{Berger} but uses tapering at coarse-fine 
interfaces~\cite{Lehner:2005vc}.

Among the publicly available MPI based AMR toolkits,
several have demonstrated weak scaling to thousands of processors ~\cite{Loeffler,Straalen,Teyssier}.
Strong scaling from one or very few processors up to a large numbers of processors,
however, has proven to be much more difficult to
achieve ~\cite{Wissink,Wissink2,Teyssier,Luitjens,Anderson}.  The generic lack of robust
strong scaling in the available MPI based AMR toolkits frequently leaves researchers
relying on data checkpointing techniques for long periods of time because they cannot
effectively utilize more than a few hundred processors on a machine with tens of
thousands of available processors. 
The ParalleX execution model aims to improve strong scaling in scaling impaired algorithms such as AMR
by providing the semantic constructs which can remove global timestep barriers.

\begin{figure}[htp]\centering
\includegraphics[scale=0.62]{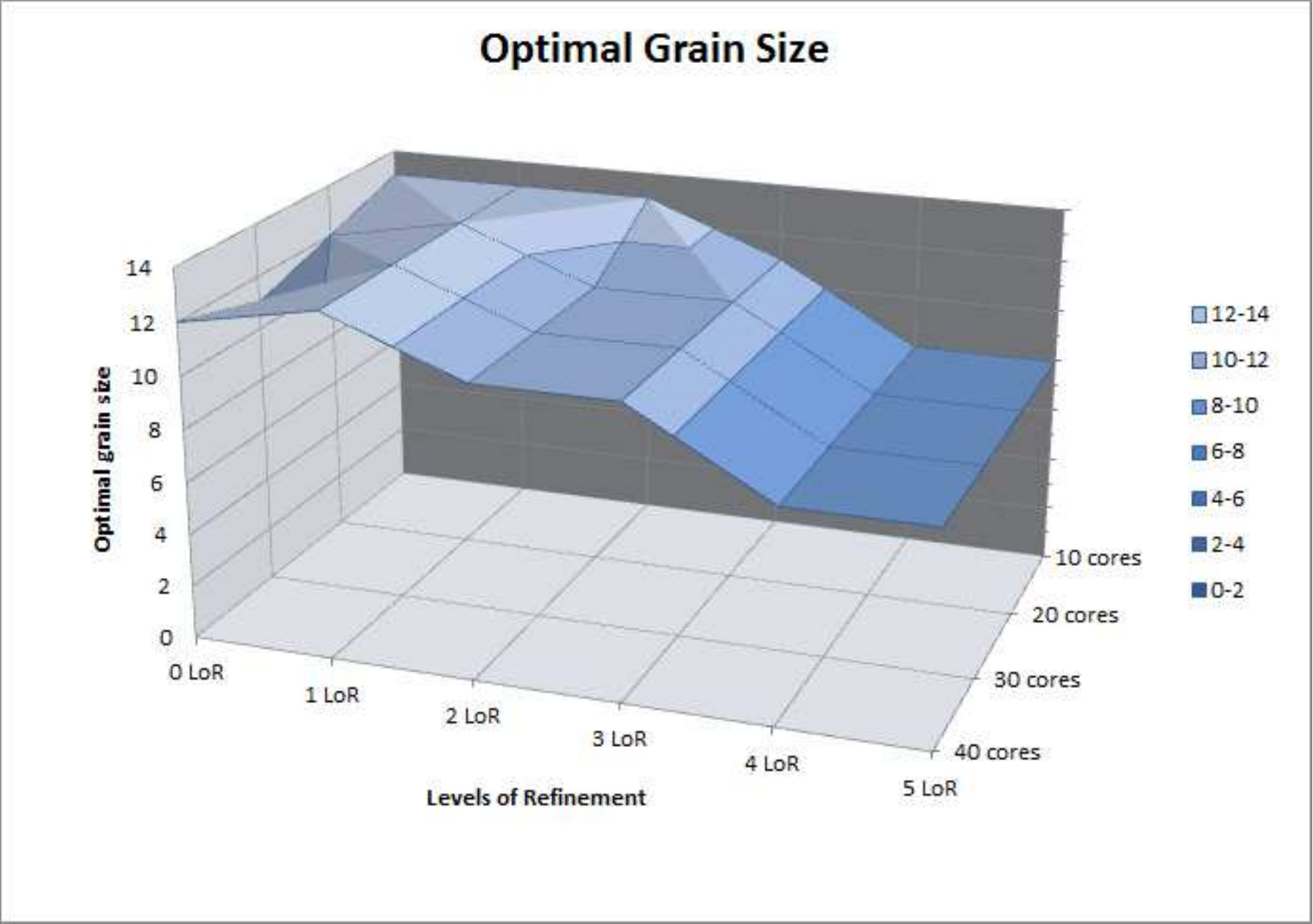}
\caption{This figure shows the optimal task granularity (or grain size) 
for a ParalleX based mesh refinement 
simulation in 3-D solving the homogeneous version of Eqns.~\ref{eq:chi}--\ref{eq:critical} as a function of number of 
levels of refinement and number of cores.  This plot was produced experimentally.
The optimal grain size does not seem to depend heavily 
on the number of cores requested.}
\label{fig:optimal}
\end{figure}

\begin{figure}
\begin{tabular}{cc}
\epsfig{file=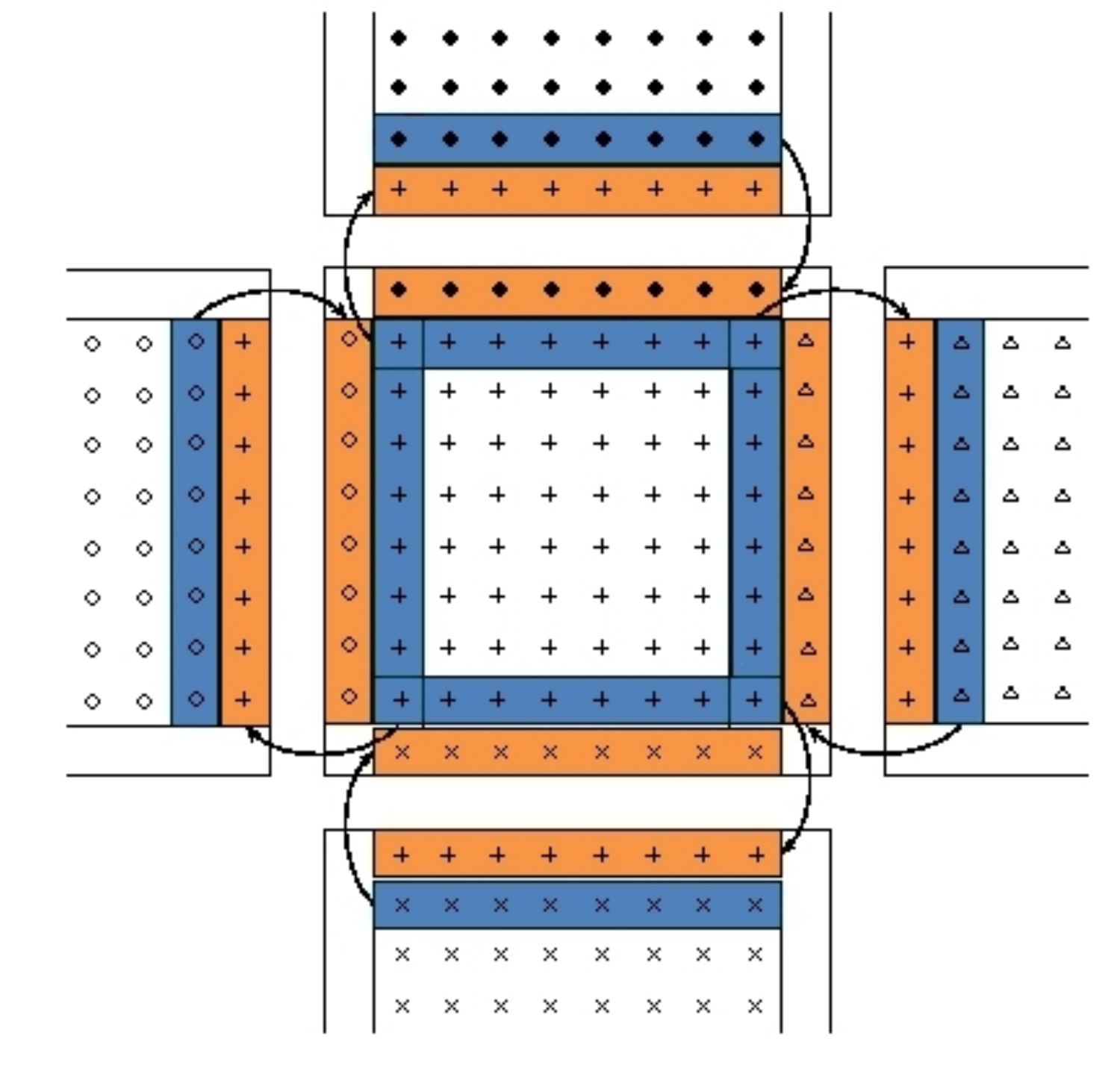,width=.26\textwidth} & %,height=7.5cm} &
\epsfig{file=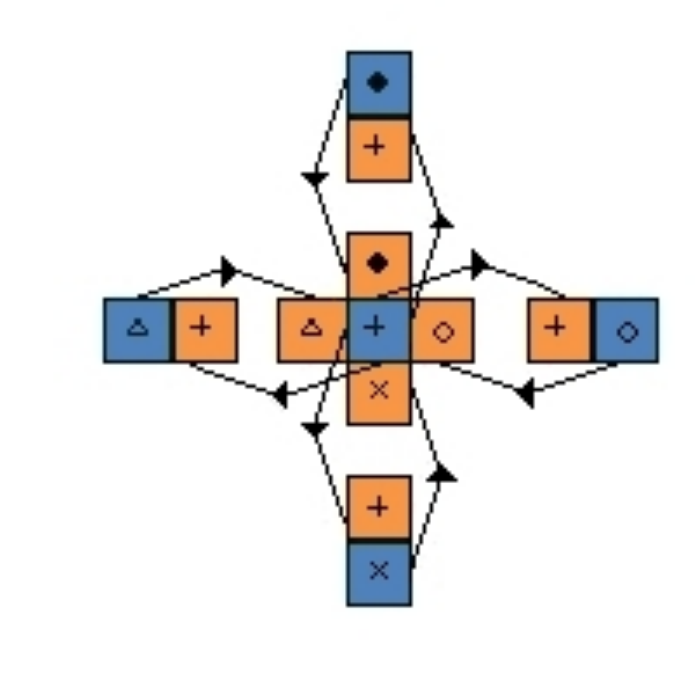,width=.19\textwidth} \\ %,height=5.0cm} \\
{\bf (a)} & {\bf (b)}
\end{tabular}
\caption{\small{Two different approaches to structured mesh based communication: (a) 2-D representation of a typical communication pattern for a finite difference based AMR data block.  Large blocks of memory are passed to the user for computation where only the boundaries of the blocks are communicated among processors. The orange regions (ghostzones) are communicated regions originating from blue zones on a distributed memory block.  The extreme limit of this communication
pattern is seen in (b): each point in the computational domain is communicated.
ParalleX based AMR is capable of smoothly transitioning between both paradigm (a) and (b) 
by means of a runtime parameter so 
that the user can adjust the optimal task granularity for a particular simulation configuration.
}} \label{fig:granularity}
\end{figure}

We note that the HPX implementation of the ParalleX model is also capable of implementing the 
standard AMR algorithm with global barriers
as it is typically implemented when using MPI.
However, HPX provides semantic constructs enabling the user to eliminate global timestep barriers while still
respecting the causality of the algorithm thereby giving the simulation greater flexibility in the order of 
computation.  We follow this latter approach.

The ParalleX based AMR we present here removes all global barriers to computation, including the timestep barrier, and 
enables autonomous adaptive refinement of regions as small as a single point without requiring knowledge of 
the refinement characteristics of the rest of the grid.  The hyperbolic partial differential equation
solved in this application ensures that the domain of dependence of each point is much smaller
than the global computational domain.  Incorporating the domain of dependence 
into the dataflow LCO construct gives greater flexibility as to when the timestep for a particular point is 
updated: points in the computational domain are updated when those points in their domain of dependence have 
been updated.

With the global timestep barrier removed by using dataflow LCOs, the application code can adjust the 
desired task granularity as a parameter in order to optimize for a specific simulation configuration.  
The optimal task granularity will vary according to the hardware architecture, problem size, and even the number of
processors requested.  Figure~\ref{fig:optimal} shows the 
optimal task granularity (or grain size) for 
the homogeneous version of Eqns.~\ref{eq:chi}--\ref{eq:critical} using mesh refinement in 3-D.    

Finite difference AMR codes typically select task granularity determined by clustering algorithm requirements.
Clustering algorithms pick the largest task granularity possible in order to reduce overhead; these
large memory blocks of grid points are passed to the user defined code and only the boundaries 
of these blocks are 
communicated between processors as illustrated in Fig.~\ref{fig:granularity}(a).  
In the ParalleX based AMR code explored here the user selects the task granularity.  The task granularity
can even be as small as a single point (See Fig.~\ref{fig:granularity}(b)). 
In a work queue based execution model, the optimal task granularity may be much smaller than that
suggested by a clustering algorithm.  We also find this to be the case with ParalleX (See Fig.~\ref{fig:optimal}).
The fine grain task granularity capability of ParalleX coupled with the timestep update flexibility 
provided by the dataflow LCO construct distinguish this ParalleX based AMR code from traditional MPI based AMR codes.

\section{Analysis}
\label{sec:analysis}

In this section we explore performance and scaling results from the ParalleX based AMR implementation.  
We observe that implicit load-balancing occurs in parallel AMR simulations as a result of the
message-driven work-queue execution.  We compare performance results to an
MPI code executing the same numerical relativity application and using the same AMR grid structure.
We explore the overhead and scaling associated with the lightweight HPX-threads used in the 
HPX implementation.  
We also demonstrate that the prototype HPX implementation
is capable of
implementing highly nontrivial AMR problems,
running in parallel across an order of magnitude of processors with 
extremely fine task granularity, 
and implementing parallel memory management for asynchronously accessed dynamically refined mesh objects. 

Data presented in this section was generated using either one of two
1$+$1 AMR codes: the HPX based code, \had\,\, or its MPI based counterpart.
Both of these codes solve the semilinear wave equation with 
exponent $p=7$ with second order finite differencing in space and 
Runge-Kutta third order integration in time.  

\begin{figure}
\epsfig{file=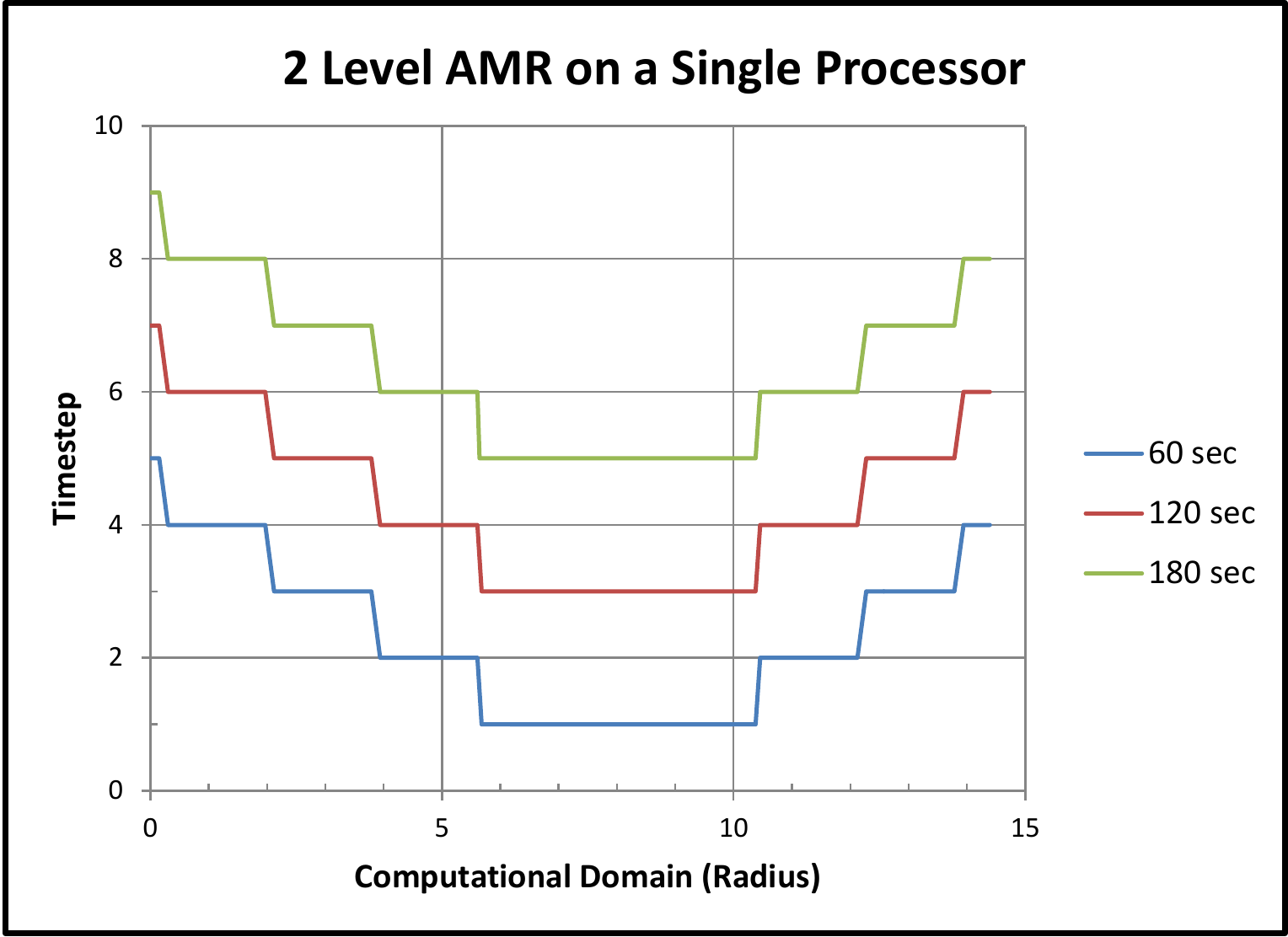, width=0.95\linewidth} %,height=9.5cm,angle=270}
\caption{\small{Snapshots at various wall clock time intervals of the timestep each point in the computational domain has reached; when global barriers are removed, some points in the computational domain can proceed to compute more timesteps than others in a fixed amount of wall clock time.  This is a simple consequence of replacing global barriers with point-to-point synchronization.  
The dataflow LCO construction ensures that causality is still respected in the computation.  Consequently the timestep curve takes on an upward facing cone shape.    
Note that the user still has to wait until all timesteps are complete in order to use the result.  However, the independence of different regions of the computational domain gives more flexibility in order to better load balance the AMR problem. Consequently, when comparing parallel AMR simulations with and without global barriers, the case without global barriers simulates faster than the case with global barriers because of better load-balancing.
}} \label{fig:noglobalbarrier}
\end{figure}

\begin{figure}
\centering
\begin{tabular}{c}
\epsfig{file=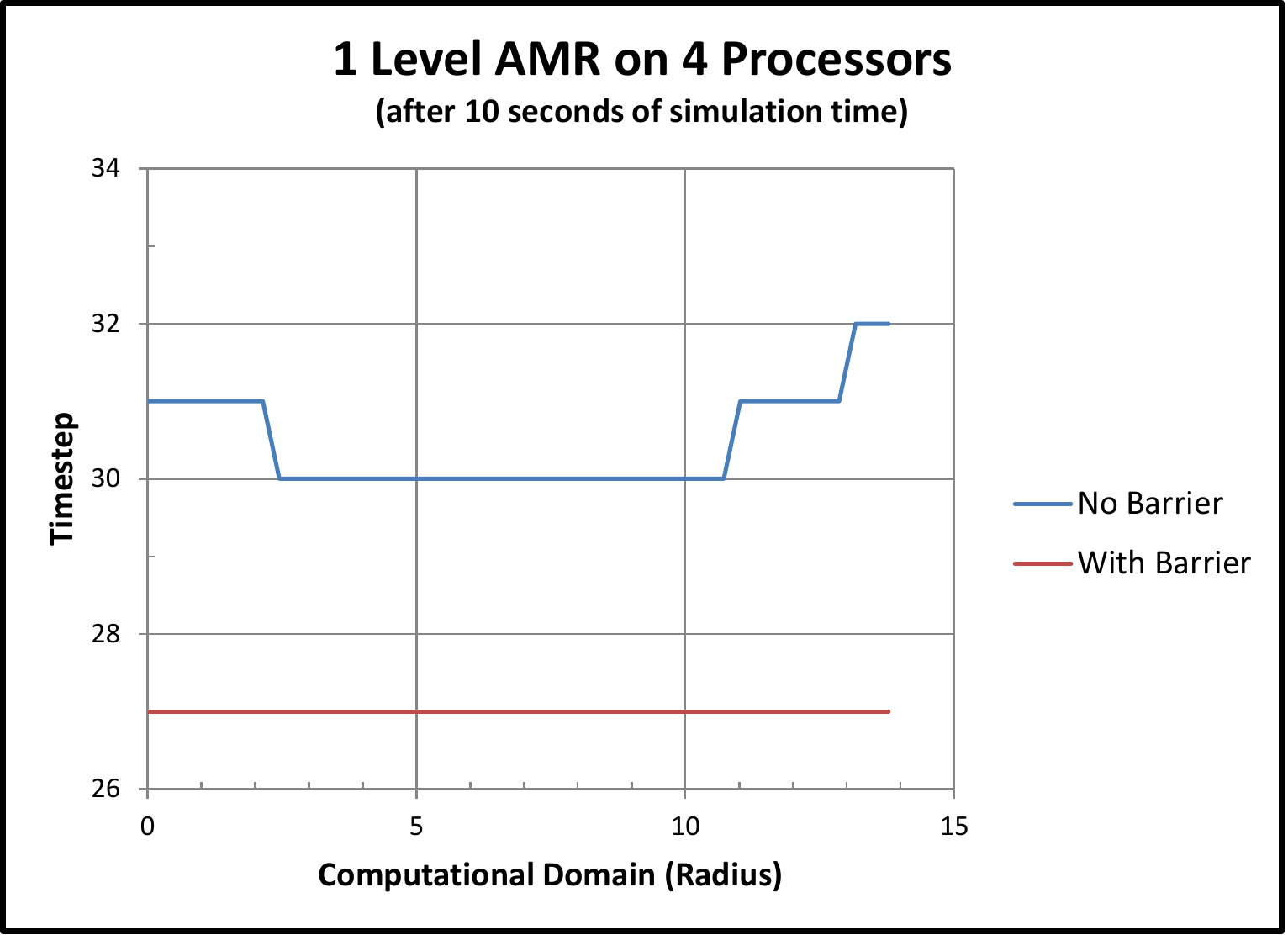, width=0.95\linewidth} \\
{\bf (a)}  \\
\epsfig{file=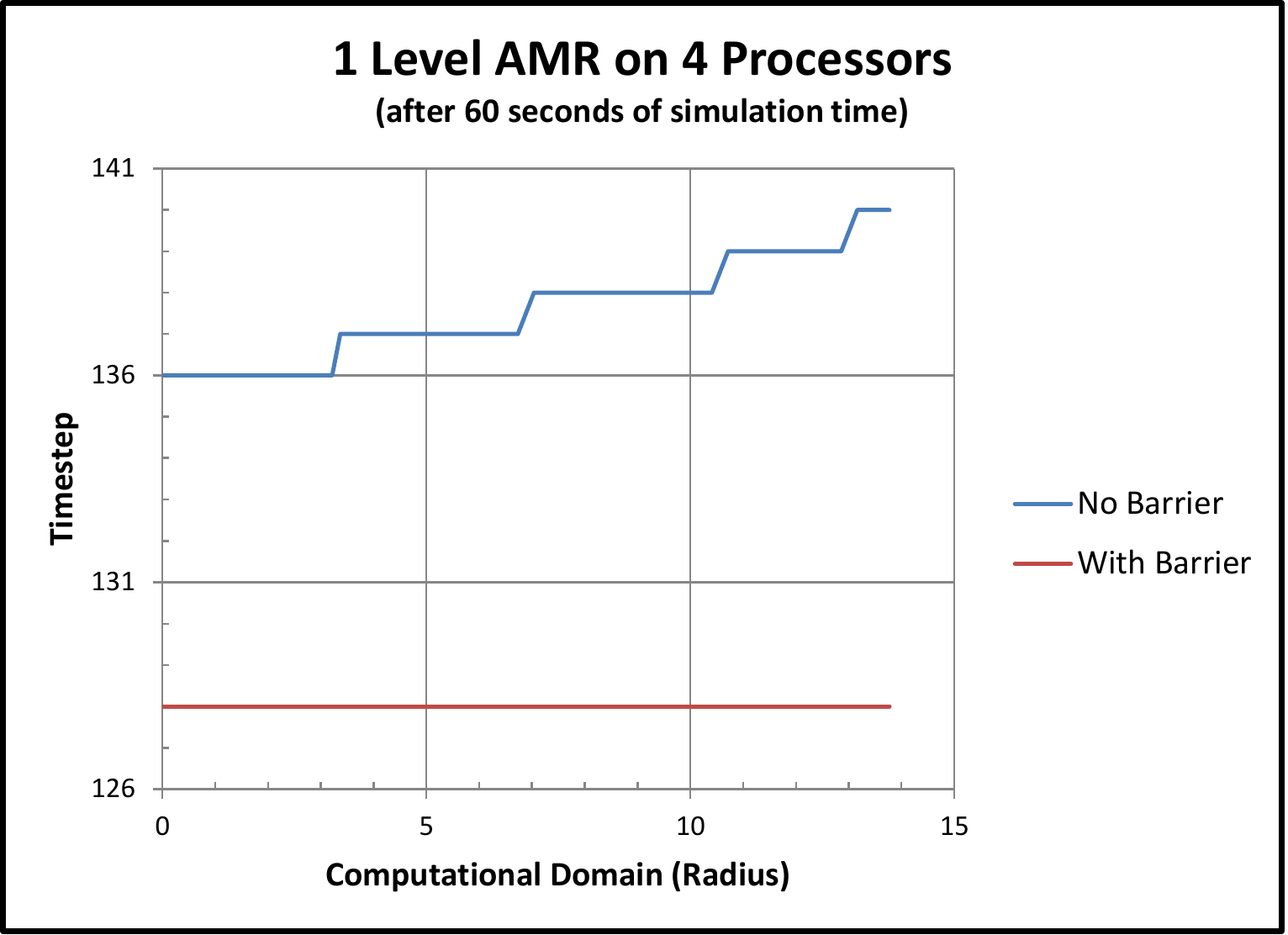, width=0.95\linewidth} \\
{\bf (b)} \\
\end{tabular}
\caption{\small{Illustration of the impact of implicit load balancing.  This plot compares the timestep reached by every point in the computational domain after either 10 or 60 seconds of wall clock time for an AMR simulation with 1 level of refinement.  The refinement criterion was scalar field amplitude. (a) and (b) show results performed on four processors. Removing the global timestep barrier gives more flexibility to load balance; consequently the parallel cases which don't enforce a global barrier are able to compute faster than those cases in (a) and (b) which do enforce a global timestep barrier.}} \label{fig:1levelamr2}
\end{figure}

Some points in the computational domain of an AMR simulation compute faster than others.  If a global timestep
barrier were in place, all points in the computational domain would have to wait for the slowest point 
in the domain to update before proceeding to compute the next timestep.  
With ParalleX based AMR, tasks proceed once the points in their domain
of dependence are available. 
Figs.~\ref{fig:noglobalbarrier} and ~\ref{fig:1levelamr2} illustrate the impact of this.  Higher resolution mesh
points take longer to compute than coarser resolution points.  These coarser mesh points are often able to compute
several timesteps ahead of their finer mesh point counterparts instead of waiting for those finer mesh points
to compute before proceeding.    
When computing on just one processor, removing the timestep barrier has no performance impact on a simulation.
But when computing on several processors, the thread task manager acts as load balancer
ensuring that processors have a steady stream of tasks and a faster total execution time results. 
As the task granularity becomes finer, the process becomes even more efficient:  processors spend less time waiting
for work to become available to them than in larger granularity counterparts.

Fig.~\ref{fig:noglobalbarrier} is a 2 level AMR simulation, or a simulation with three different resolution meshes 
-- a coarse mesh, a finer mesh, and a finest mesh.  The singularity threshold formation search was run for 
60, 120, and 180 seconds of wall clock time.  The timestep that each point in the spatial computational domain 
had reached by the end of the 60, 120, or 180 seconds is plotted.  Unlike in MPI simulations with global barriers, 
these simulations show that some points in the computational domain are able to compute several 
more timesteps than others in the same amount of wallclock time.  
Futures ensure that causality is respected by requiring that the immediate neighbors of a point being updated 
be at the same timestep as the point being updated.  
Thus the resulting timestep curve in Figs.~\ref{fig:noglobalbarrier} and ~\ref{fig:1levelamr2} 
resembles an upward facing cone where the tip of the cone is located in the region of highest spatial resolution.

In Figure~\ref{fig:1levelamr2}, we compare AMR simulations with 1 level of refinement running with and without a global timestep barrier on four processors.  AMR simulations were run for either 10 or 60 seconds of wall clock time and the timestep reached by each point in the computational domain was plotted.  Cases without the global barrier were able to compute more timesteps than cases with the global barrier in the same amount of time.  This is a natural consequence of the message-driven work-queue execution model: processors are able to overlap communication and computation more efficiently than algorithms which enforce a global barrier every timestep.

\begin{figure}
\begin{tabular}{c}
\epsfig{file=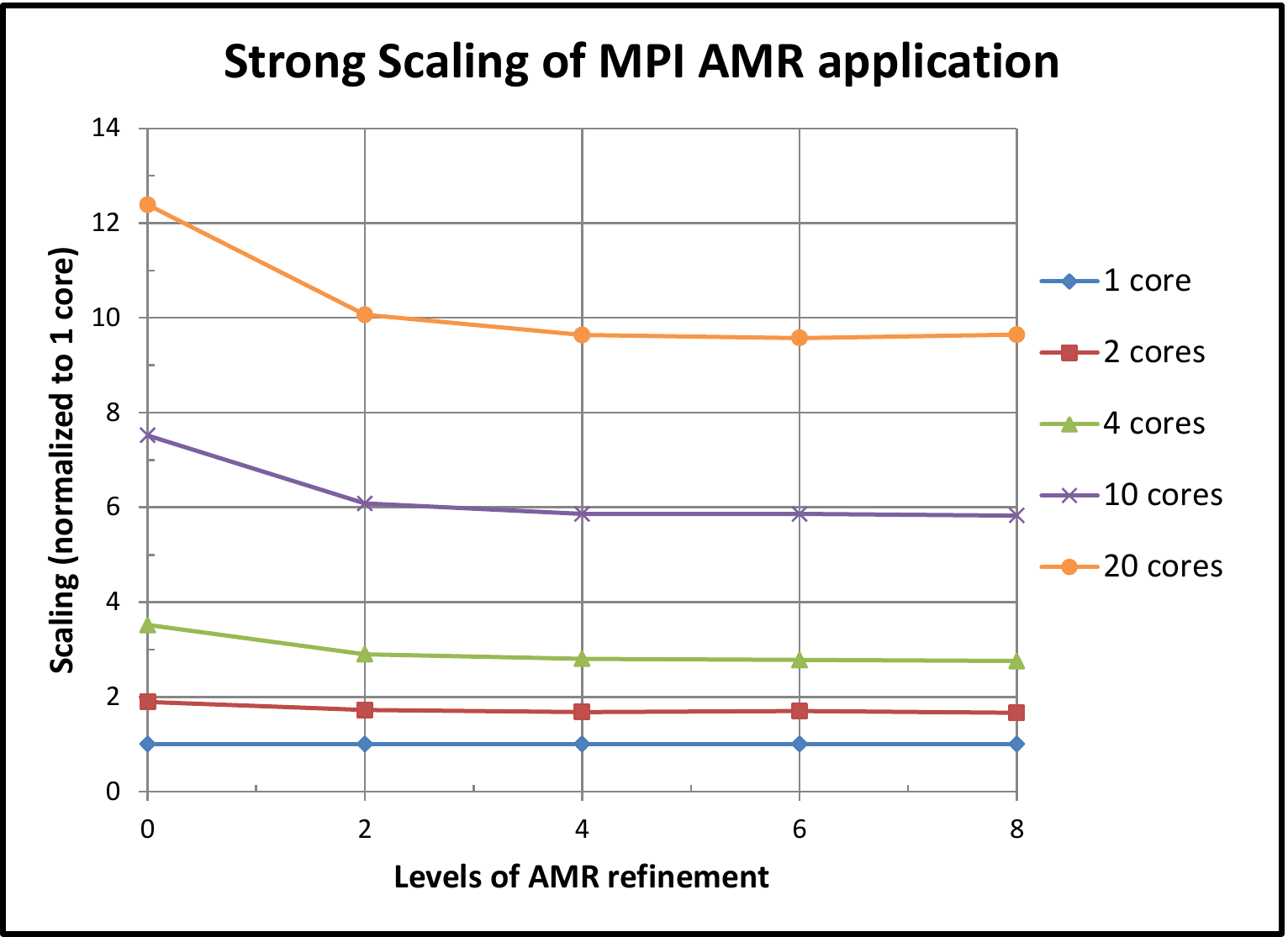, width=0.95\linewidth} \\
{\bf (a) }  \\
 \epsfig{file=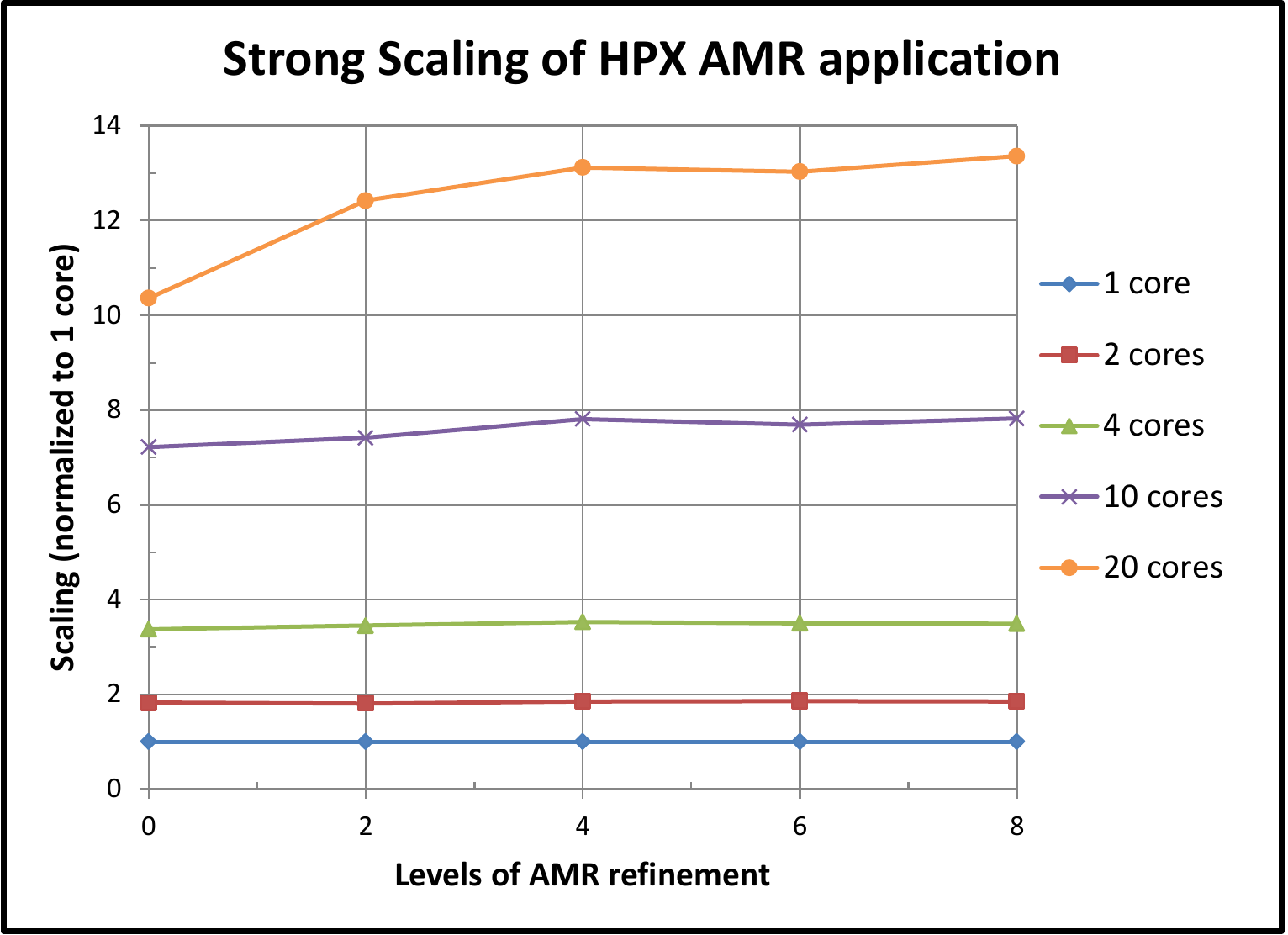, width=0.95\linewidth} \\
{\bf (b) } \\
\end{tabular}
  \caption{\small{{\bf (a)} Strong scaling results using MPI-based mesh refinement.
 The vertical axis
shows the speedup of the application compared to running on a single core.
 As indicated in the horizontal axis, various numbers of refinement levels were tested.
For cases run on more than 1 core, the plots are monotonically decreasing.  This illustrates
 a
well-known result that MPI mesh refinement applications show worse strong-scaling
as the number of levels of refinement in the simulation is increased. {\bf (b)}
Strong scaling results using ParalleX-based mesh refinement.  The strong scaling improves as the number
of levels of refinement increases.  In terms of wallclock time, the ParalleX
based code begins to outperform the MPI version whenever the simulation has more
than four levels of
refinement and is run on 10 cores or more.  See Fig.~\ref{fig:mpi_hpx_wallclock}.}
  }
\label{fig:px_scale}
\end{figure}

\begin{figure} \centering
\includegraphics[width=0.95\linewidth]{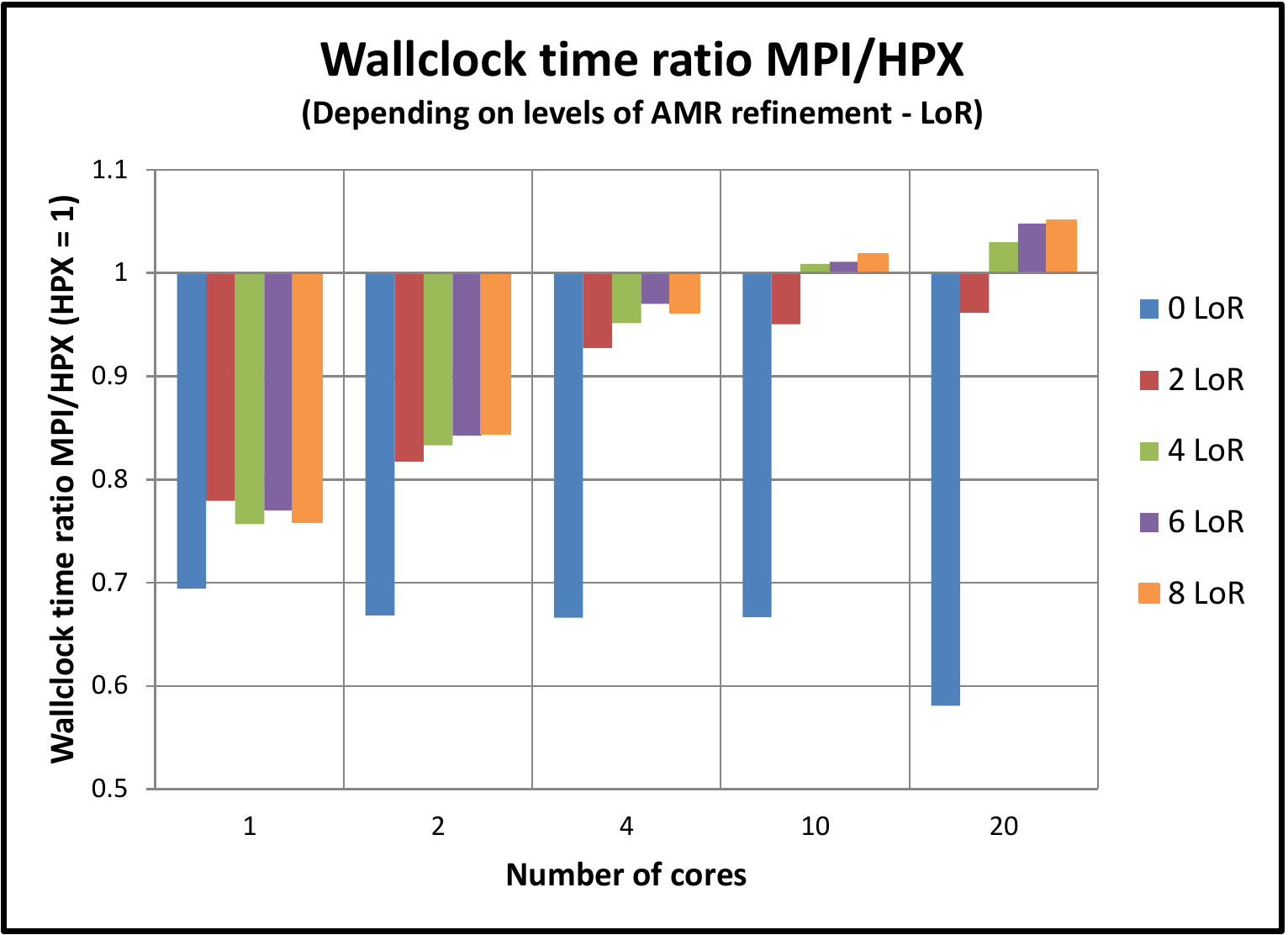}
  \caption{\small{Wallclock time performance comparison between the MPI based and HPX
based mesh refinement.  The HPX based code adds overhead
compared to its MPI counterpart which results in slower execution in simulations with fewer
levels of refinement.  MPI outperforms HPX in these cases.  However, as the number of levels
of refinement increases and as the number of processors increases, the HPX code
outperforms the MPI counterpart by as much as 5\%.  While HPX adds overhead
compared to MPI, it also reduces starvation and latency compared to MPI.}
  }
\label{fig:mpi_hpx_wallclock}
\end{figure}

\begin{figure} \centering
\epsfig{file=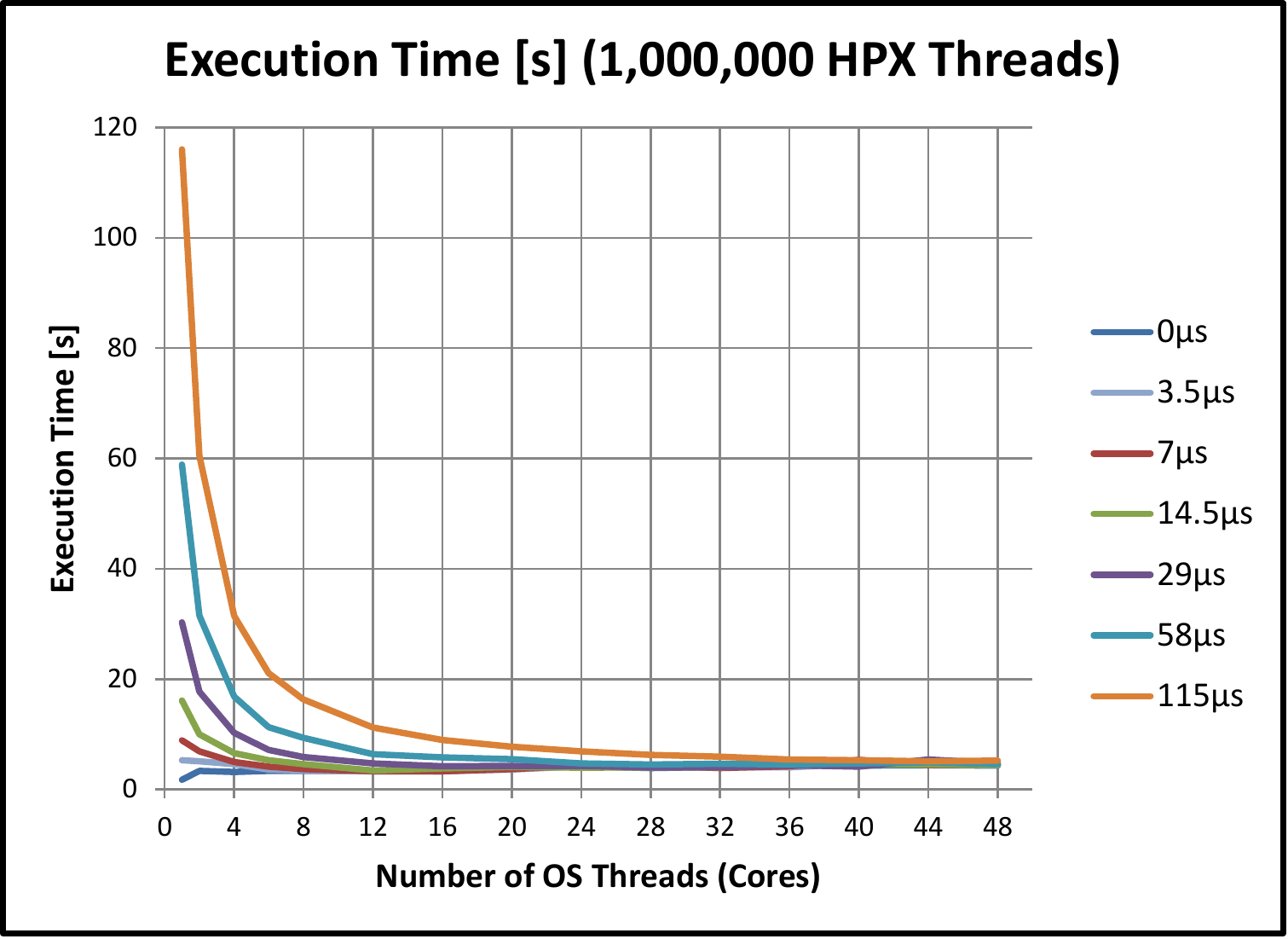, width=0.95\linewidth} %,height=10.0cm,angle=270}
\caption{\small{
The HPX runtime system
exhibits adequate scaling levels when measured by the average overhead
of HPX-thread management on an SMP machine.  The number of cores utilized is controlled
by the number of operating system threads used and is increased from 2 to 48 cores.
Each of colored lines represent a different amount of artificial workload executed by each
thread.  The lowest blue line executed no workload, all the time is overhead and so there
is no scaling.  At the other end of the workload spectrum, the top orange line adds
115 $\mu$s of wait time on each thread; in this case, the average overhead over
the one million threads is reasonable and a fair scaling factor of almost 23 is achieved when
running on 44 cores.}
}
\label{fig:thread_execution_time}
\end{figure}

Scaling and performance comparisons between HPX and MPI based AMR are shown in
Figs.~\ref{fig:px_scale} and~\ref{fig:mpi_hpx_wallclock}. 
In application performance experiments, the HPX runtime system substantially
reduced starvation and latency effects which resulted in better load-balancing and better strong scaling
than comparison code written using MPI.
As levels of refinement were added to the simulation, strong
scaling {\it improved} in the HPX version.  The MPI comparison code showed the opposite behavior:
strong scaling decreased as levels of refinement were added.
The reduction in starvation and the mitigation of latencies when using the HPX runtime 
system comes at a cost of increased overhead and contention.  
Some of this overhead can be controlled and partially amortized by adjusting the
task granularity of a simulation.  This reduces the number of lightweight threads used and allows the user
to optimize the granularity for a particular simulation configuration.   Thread
overhead is $\sim$3-5 microseconds (see Fig.~\ref{fig:thread_execution_time}).  
Simulations often use as many as $10^8$ threads or more.

These results suggest that the ParalleX execution model can substantially improve the scalability in 
scaling impaired applications by reducing starvation and latency.  But there is clearly more
overhead costs associated with this execution model.  Applications employing very regular datastructures
with reliable, near-identical workloads for each component will perform better using the CSP model than
ParalleX due to lower overhead.  Such applications already scale extremely well using the CSP model.

Work on developing a theoretical model for ParalleX
which quantitatively details the trade-offs between improved scalability and higher overhead is
currently underway.  Some of this overhead can be amortized through task granularity control; however, 
a key component of a successful implementation of the ParalleX execution model 
includes hardware acceleration of the runtime system.  We address this topic in the following section.

%%%%%%%%%%%%%%%%%%%%%%%%%%%%%%%%%%%%%%%%%%%%%%%%%%%%%%%%%%%%%%%%%%%%
\section{Runtime System Acceleration}
\label{sec:implications}

Performance analysis of various HPX applications indicates that many runtime
system constructs are a significant source of overheads when implemented
purely in software, or when utilizing operating system primitives forcing
context switches.  We explore the possibility to alleviate these overheads through offloading a number of
frequently used system functions to Field-Programmable Gate Array (FPGA) technology.
The benefits of such an approach are threefold: firstly, the code related
directly to the computation will not be disturbed by ``housekeeping'' duties,
potentially avoiding unnecessary invalidations of cache contents, translation 
lookaside buffer (TLB)
thrashing, and eliminating wasteful switches to kernel mode.
Secondly, many modern processors, despite implementing extended instruction
sets, still exhibit poor performance levels for certain code types
(bitwise searches, bit counting, pattern matching and correlation finding).
Thirdly, the mechanisms used to synchronize thread execution on multiple cores
use rather simplistic memory-oriented abstractions, offering little in the way
of supporting more sophisticated constructs.
Finally, the experiences gathered during the experiments with the FPGA
based accelerators along with the developed hardware description language (HDL) code base
will serve as a foundation for future in-silicon implementation of 
the high-performance accelerator logic to be integrated with 
existing systems, or to provide building blocks of next generation
processors.

Below we discuss the five candidate system software areas we have identified for which the
implementation of supporting functions in hardware is likely to bring
the most pronounced performance gains for the ParalleX execution model.

\B{Atomics:}
The atomic synchronization functions available in libraries
providing the operating system (OS) thread application programming interface (API) 
often incur significant performance penalties.
The most substantial costs are associated with switching into kernel space
when contention (simultaneous access from more than one thread) is
detected.
Even when no contention occurs, the user space wrapper code uses
high-overhead instructions resulting in locking the memory bus, or, at
the very least, restricting the concurrent accesses to the affected cache
lines.
We are investigating an FPGA-managed
implementation of atomic functions, with primary focus on developing
an efficient support for ParalleX LCOs.
We expect that while the latency of communication with the FPGA will
be higher than that of an average scalar memory access, the synergy between
the following implementation properties will improve the overall performance
of LCO functions:
\begin{enumerate}[(a)]
\item automatically enforced serialization of communication packets
(by the north bridge/motherboard chipset);
note that there still may be multiple ``concurrent'' requests serviced by
the FPGA due to unpredictable packet fragmentation induced by
some distributed bus protocols (e.g., PCI-Express),
\item thread safety achieved by assigning individual hardware request buffers
with associated control logic to every OS thread in active use,
\item enforcement of predefined access order, which could be based, for
example, on relative thread priorities,
\item background main memory updates performed by HyperTransport, QuickPath 
or PCIe devices
modifying the state of atomic objects independently from the processor;
the correctly defined producer-consumer relationships between the central
processing unit (CPU) and
FPGA become critical to avoid the reintroduction of race conditions, and
\item atomic object state offload to dedicated memory pools attached to,
or on the FPGA devices, leaving only a portion of the LCO that is directly
relevant to the computation exposed to the processor and placing the
associated control state under the exclusive control of the FPGA.
\end{enumerate}

We are investigating atomic support for the full range of ParalleX LCOs,
including lightweight LCOs such as {\it mutex} and {\it counting
semaphore}, which mimic typical synchronization primitives found
in thread programming libraries, and high-level LCOs with more 
sophisticated semantics, of which the most prominent examples are
{\it dataflow} and {\it future}.
We expect that the experiences gathered during their implementation will
shed more light on trade-offs involved in hardware acceleration of generic
synchronization mechanisms.

\B{Memory management:}
Dynamic memory allocation is an indispensable mechanism for frugal
management of storage resources required for temporary objects and
data structures, frequently generated by higher level programming
languages.
However, due to the complexity of dealing with fragmentation and
random request sizes, the task of managing the content allocation
within the virtual arena is typically left to the runtime software,
thus introducing undesirable latencies in program execution.
During the course of HPX development we observed that such overheads can
be noticeable even for arrays of similarly-sized objects, such as
user space threads.
Replacing or augmenting software allocators would provide immediate
benefits by shifting the substantial portion of management tasks to the
background, where they may be performed concurrently with the computation.
While we don't plan to replace the OS management of physical memory pages,
FPGAs will be used to handle preallocated virtual memory arenas.
In the simplest case, they will apply bit-mapped allocators for pools
of uniform objects; since finding the next available slot can be performed
eagerly, most allocation requests might be satisfied immediately.
Furthermore, the FPGA control logic may be augmented with heuristic grouping
objects known to be subjected to similar access pattern into the smallest
number of memory pages, thus bringing down the TLB miss rates.
Based on the results of investigations, 
we anticipate extending the hardware allocator to support a
broader range of request types along with an improved heuristic to
mitigate the unwanted system phenomena (cache and TLB misses, mapping to
remote memory pages on SMP platforms, etc.).
Finally, combining allocation with copying of memory contents involving a
hardware direct memory access (DMA) engine (as opposed to CPU data pipeline) can expedite the
execution of such functions as {\tt realloc} or simple C++ copy
constructors.

\B{Thread scheduling:}
A well optimized user-level thread scheduler is fundamental to an
efficient ParalleX runtime implementation.
Even though the crossing of the kernel-user space boundary is mostly
avoided in this case, there are other issues impeding the performance
of the software-only implementation, increasing the complexity of its
design.
First, the user thread queue(s), irrespective of whether a single or
multiple instances were created per locality,
% by the runtime system, 
represent
a shared resource accessed by multiple operating system threads.
Thread insertion and dequeue operations, preferably supported on both
ends of the queue to facilitate work stealing, must therefore be atomic.
Second, the scheduler must act in accordance with thread priorities and
other parameters imposed by scheduling policies.
Third, context switching (processor register state save and restore)
generates multiple accesses to the stack resulting in relatively
high execution latency even if written in assembly language.
Finally, since the scheduler needs to deal with the creation and termination
of a potentially high quantity of ephemeral threads, seamless cooperation
with thread spawning and killing entities (parcel handler,
other threads) is highly desirable.

The FPGA scheduler must address most of these issues, hence it is more
involved than being just a collection of priority queues with atomic
access.
These investigations will determine the trade-offs applicable to the hardware
scheduling support, which aspects of the decision logic can be actually
transplanted from the software world onto an FPGA, and how to distribute,
and if necessary, replicate efficiently the thread state data across the
main memory and FPGA registers (or other topologically close memory
resources).
Finally, since current mainstream processors don't incorporate any
lightweight mechanisms allowing them to react to external events (other
than using expensive interrupts), we are searching for the best approach
to minimize latencies of control transfer between the FPGA and CPU.
We also anticipate that the perfected hardware FIFO implementation
may find applications outside thread scheduling.

\B{Parcel-driven operations:}
Active message processing requires a tight integration of system-level
interconnect channels with local execution resources, which traditionally
has been an Achilles-heel of software implementations.
Many such systems use the kernel as an intermediary in passing the data
between the network hardware and the applications, paying the cost
of extraneous context switches, multiple message buffers, and increased
latencies and/or resource usage to arrange the timely instantiation of
threads triggered by message arrival.
We propose to combine the idea of ``intelligent'' network 
interface controller (NIC)~\cite{druschel_nic_1994}
with FPGA logic to minimize the reliance on in-kernel message processing
and enable a streamlined handoff of data and control flow to the
application user space.
Hardware support for the parcel handler includes the ability to deal
with pure data transfers (from single scalars through the serialized
ParalleX objects), simple atomic operations on memory
(AMOs) that can proceed without the involvement of the processor, LCO
state updates (compound memory operations), and instantiation of user
threads to execute remote actions.
The FPGA design must be able to cooperate with both software and
hardware-level thread schedulers, albeit possibly achieving different
performance levels.
FPGA boards integrating a 1~Gbps TEMAC (Tri-mode
Ethernet Media Access Controller) chip connected to standard RJ-45 socket
or SFP (Small Form-factor Pluggable) connector interfacing directly to the
FPGA pins (available from multiple vendors) are a particularly useful 
research vessel in this endeavor.

\B{AGAS:}
Hardware mechanisms that aid object namespace management hold an additional
promise of eliminating much of the computational overhead of object
lookups.
Fully associative approaches similar to those utilized by
TLBs may be difficult to apply due to the sheer number of object names 
(on the order of tens of thousands and more) maintained per locality.
However, a form of hashing, perhaps based on simplified cryptographic 
algorithms, may provide a lightweight and energy-efficient mechanism
to locate the physical address translation for a given global object name.
A further integration with parcel layer may result in shortened and
highly efficient path for processing of those types of received 
parcels that do not require explicit use of software threads, such as AMOs.

To evaluate the viability of offloading the system functions to hardware,
an early implementation of a global thread scheduler queue was developed in
Verilog.
The resulting logic configuration was uploded to a Xilinx Virtex-5 FPGA 
on a 4-lane PCI-Express board clocked at 125 MHz. 
The hardware-augmented implementation was able to match and in most cases 
marginally surpass the performance of an eqivalent software only queue 
on a thread-intensive Fibonacci benchmark.
These early results are encouraging, since the hardware implementation 
did not support efficient DMA operations and was tested with a generic 
PCI connectivity library ({\it libpciaccess}) instead of a properly tuned 
kernel driver.
Analysis of internal timings collected with logic analyzer soft cores
({\it Chipscope}) revealed that all PCI read requests issued by the 
application were unnecessarily limited to payload sizes of at most 4 bytes,
effectively adding the latency of roughly 90 FPGA cycles, or 720 ns, per 
request, which is several times greather than the average memory access
cycle on our test platform.
It is expected that addressing these inefficiencies will result in a
significant performance boost of hardware functions.

In future work, we will expand and optimize these results with additional 
custom hardware modules developed to accelerate all five system functions.   
While FPGAs have been in existence for well over
two decades and have been applied in accelerating various computational algorithms in 
hardware, the approach we explore is fundamentally different:  we explore the acceleration
of execution model system functions.

%%%%%%%%%%%%%%%%%%%%%%%%%%%%%%%%%%%%%%%%%%%%%%%%%%%%%%%%%%%%%%%%%%%%
%
%   C O N C L U S I O N S
%
%%%%%%%%%%%%%%%%%%%%%%%%%%%%%%%%%%%%%%%%%%%%%%%%%%%%%%%%%%%%%%%%%%%%
\section{Conclusions}
\label{sec:conlusions}

We have presented the ParalleX execution model and the HPX runtime system implementation, 
a distributed parallel AMR application framework which employs dataflow LCOs to eliminate global timestep barriers, 
performance results using HPX for a numerical relativity application,
and a roadmap into hardware acceleration of system functions in the execution model itself.
We find that the ParalleX execution model is capable of reducing starvation and latency in complex applications
at a cost of higher overhead.  We have demonstrated how ParalleX is capable of expressing finer-grained 
dependencies than MPI and thereby eliminates global barriers.  The cost of expressing these finer-grained
dependencies is higher overhead and the use of more lightweight HPX threads in accomplishing the same amount work.  
When the proper balance between these competing factors was reached,
we found that the HPX based AMR code can both outscale and outperform the MPI based AMR code. 

  In spite of the additional overhead that HPX requires, we found that the software-only 
implementation of the HPX is still capable of both outscaling and outperforming 
the MPI AMR code implemented with global barriers.  We present evidence that FPGAs can further reduce
this overhead and form an integral part in implementing the execution model.
We note that the HPX implementation of ParalleX is freely available under an open source license
along with many example codes, including the test codes used for this paper \cite{tutorial}.
The ParalleX execution model offers the potential of a viable path to the Exascale systems of the future and 
the scaling impaired applications of today.

%%%%%%%%%%%%%%%%%%%%%%%%%%%%%%%%%%%%%%%%%%%%%%%%%%%%%%%%%%%%%%%%%%%%
%
%   A C K N O W L E D G M E N T S
%
%%%%%%%%%%%%%%%%%%%%%%%%%%%%%%%%%%%%%%%%%%%%%%%%%%%%%%%%%%%%%%%%%%%%
\noindent{\bf{\em Acknowledgments:}}
It is a pleasure to thank Steven L. Liebling and Steven Brandt for useful discussions and comments.  We
acknowledge support comes from NSF grants 1048019 and 1029161 to Louisiana State University.

\nocite{*}

\bibliographystyle{apsrev}
\bibliography{pxBib}

\end{document}